\documentclass[cameraready]{Interspeech}

\title{Beyond Cross-Reconstruction: Probing-Based Disentanglement Evaluation \\ for Acoustic Teleportation Codecs}

\author[affiliation={1}]{Philipp}{Grundhuber}
\author[affiliation={2}, orcid=0000-0002-2613-8046]{Emanu{\"e}l A. P. }{Habets}

\address{
$^1$ Fraunhofer Institute for Integrated Circuits (IIS), Erlangen, Germany \\ 
$^2$ International Audio Laboratories Erlangen\textsuperscript{$\ast$}, Erlangen, Germany \thanks{\textsuperscript{$\ast$}A joint institution of Fraunhofer IIS and Friedrich-Alexander-Universit{\"a}t Erlangen-N{\"u}rnberg (FAU), Germany.}}

\email{philipp.grundhuber@iis.fraunhofer.de, emanuel.habets@audiolabs-erlangen.de}


\usepackage{comment}
\usepackage{acronym}
\usepackage{pbalance}

\acrodef{AT}[AT]{Acoustic Teleportation}
\acrodef{RIR}[RIR]{room impulse response}
\acrodef{RVQ}[RVQ]{residual vector quantization}
\acrodef{NAC}[NAC]{Neural Audio Codec}
\acrodef{PCA}[PCA]{principal component analysis}
\acrodef{DNN}[DNN]{deep neural network}
\acrodef{MLP}[MLP]{multi-layer perceptron}
\acrodef{BRPE}[BRPE]{blind room parameter estimation}
\acrodef{DRR}[DRR]{direct-to-reverberant ratio}
\acrodef{RMSE}[RMSE]{Root Mean Squared Error}

\begin{document}

\maketitle

\begin{abstract}
Some neural audio codecs disentangle speech into latent subspaces encoding content, speaker identity, and acoustics, enabling acoustic teleportation and voice conversion. Existing evaluations rely on cross-reconstruction quality, which cannot reliably detect leakage across partitions. We extend a probing-based framework to assess disentanglement by regressing room-acoustic parameters (reverberation time, clarity, and direct-to-reverberant ratio) and classifying speaker identity, using the gap between intended and unintended partitions as the disentanglement measure. Applied to an acoustic teleportation codec, we find speaker identity is largely confined to its partition, while acoustics leak into the speech embeddings due to the training objective. Acoustic embeddings blindly estimate room parameters within 0.02~s of supervised baselines, indicating physically meaningful structure emerges without explicit supervision.
\end{abstract}

\section{Introduction}
\label{sec:intro}

Certain \acp{NAC} partition the latent space in order to disentangle distinct speech attributes, including linguistic content, speaker identity, prosody, and acoustic environment. This enables applications like \ac{AT}, dereverberation, and voice conversion \cite{wang2024disentangled}. Recent codecs realize this idea through diverse architectural choices: FreeCodec \cite{zheng2024freecodec} decomposes speech into global timbre, prosody, and content through separate encoders; NaturalSpeech 3 \cite{NaturalSpeech3} factors speech into five subspaces via vector quantization; SpeechTokenizer \cite{zhang2024speechtokenizer} achieves hierarchical disentanglement across \ac{RVQ} layers; SD-Codec \cite{bie2025learning} separates audio sources through parallel \ac{RVQ} branches; and Omran et al. \cite{omran2023disentangling} and Grundhuber et al. \cite{grundhuber2025acoustic} separate speech from acoustic characteristics in partitioned embeddings.

Existing approaches evaluate disentanglement through downstream task performance, such as voice conversion quality \cite{zheng2024freecodec}, TTS metrics \cite{NaturalSpeech3}, source separation metrics \cite{bie2025learning}, or cross-reconstruction after embedding swaps \cite{omran2023disentangling, grundhuber2025acoustic}. Qualitative methods (t-SNE, \ac{PCA}, listening tests) provide intuitive evidence but are difficult to compare systematically. Crucially, none can reveal whether information leaks across partitions, as a decoder may learn to ignore redundant information in the wrong partition. Classical disentanglement metrics such as DCI (disentanglement, completeness, informativeness)~\cite{eastwood2018framework}, MIG (mutual information gap)~\cite{chen2018isolating}, and the $\beta$-VAE metric \cite{burgess2018understanding} assume dimension-wise independence over discrete factors and are therefore inapplicable to the partition-level structure found in \acp{NAC}. Training lightweight classifiers on representations to evaluate disentanglement via the accuracy gap between content and speaker embeddings is an established method \cite{hsu2019disentangling, qian2022contentvec, plachouras2025unified}. Prior probing work in speech has targeted attributes such as speaker identity \cite{hsu2019disentangling, chou2019one}, but the blind estimation of continuous, physically grounded parameters based on codec representations remains unexplored.

In this work, we treat the pre-trained encoder of an \ac{AT} codec \cite{grundhuber2025acoustic}, spanning different training task sets, quantization levels, and temporal downsampling factors, as a fixed feature extractor and train identical lightweight \ac{MLP} probes on each embedding partition independently. For regression, we target the blind estimation of three room-acoustic parameters from reverberant speech (viz., reverberation time ($T_{60}$), clarity ($C_{50}$), and \ac{DRR}), which is a well-studied task \cite{lollmann2010improved, gamper2018blind}. For classification, we target speaker identity on a fixed set of known speakers. These factors serve as diagnostic probes for representation quality rather than as end goals.

Our contributions are: 1)~We introduce regression-based probing for quantifying disentanglement in \acp{NAC}, adapting the informativeness principle of the DCI framework \cite{eastwood2018framework} from individual dimensions to embedding partitions (Sec.~\ref{sec:problem}). 2)~We reveal an asymmetric disentanglement structure: speaker identity is effectively confined to the speech partition, while acoustic information partially leaks into speech partition embeddings, and trace this asymmetry to the gradient structure of the \ac{AT} training objective. 3)~We show that acoustic embeddings yield blind $T_{60}$ estimates competitive with supervised baselines, and informative $C_{50}$ and \ac{DRR} estimates, despite receiving no room parameter labels during codec training. 4)~We conduct a systematic study of the effects of training tasks, quantization levels, and temporal downsampling on disentanglement quality, demonstrating that output-quality metrics such as ScoreQ \cite{ragano2024scoreq} do not predict disentanglement and that probing is necessary to expose leakage missed by cross-reconstruction evaluation.

\begin{figure*}[h!]
    \centering
    \includegraphics[width=\textwidth]{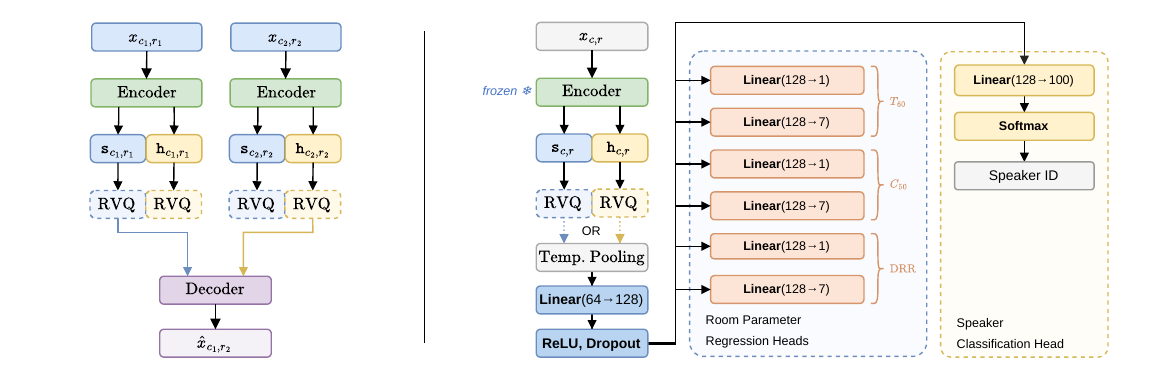}
    \caption{Probing framework for disentanglement quantification. \textit{Left:} The pre-trained \ac{AT} encoder splits reverberant speech into speech ($\mathtt{s}_{c,r}$) and acoustic ($\mathtt{h}_{c,r}$) partitions. \textit{Right:} Identical lightweight \ac{MLP} probes trained on each partition independently predict room-acoustic parameters (regression) and speaker identity (classification). The gap $\Delta_k$ (Eq.~\ref{eq:gap}) between intended and unintended partition accuracy serves as a direct disentanglement measure.}
    \label{fig:probing_framework}
\end{figure*}

\section{Problem Formulation}
\label{sec:problem}

In acoustic teleportation, a reverberant speech signal is modeled as $x_{c,r} = s_c * h_r$, where $s_c$ is anechoic speech with content~$c$, $h_r$ is the \ac{RIR} of room~$r$, and $*$ denotes convolution. The encoder splits the input into two partitions (Fig.~\ref{fig:probing_framework}):
\begin{equation}
    \{\mathtt{s}_{c,r},\ \mathtt{h}_{c,r}\} = \text{Enc}(x_{c,r}),
\end{equation}
where $\mathtt{s}_{c,r} \in \mathbb{R}^{T_s \times 64}$ captures speech content and $\mathtt{h}_{c,r} \in \mathbb{R}^{T_h \times 64}$ captures acoustic characteristics. Acoustic teleportation swaps acoustic embeddings at decode time, $\hat{x}_{c_1,r_2} = \text{Dec}(\mathtt{s}_{c_1,r_1}, \mathtt{h}_{c_2,r_2})$; dereverberation is the special case $\mathtt{h} = \mathbf{0}$.

Ideally, $\mathtt{h}_{c,r}$ encodes room-acoustics irrespective of speaker identity, and $\mathtt{s}_{c,r}$ encodes content and speaker characteristics irrespective of the room. In practice, information may leak across partitions, and such leakage is invisible to output-quality metrics like ScoreQ. We therefore seek a metric that directly measures information content per partition.

We probe each partition with three \ac{RIR} parameters defined in \cite{7336912, kuttruff2016room}:
\begin{align}
    T_{60} &: \text{time for energy decay of 60 dB}, \label{eq:T60} \\
    C_{50} &= 10\log_{10}\frac{\int_0^{50\text{ms}} h_r^2(t)dt}{\int_{50\text{ms}}^{\infty} h_r^2(t)dt}, \label{eq:C50} \\
    \text{DRR} &= 10\log_{10}\frac{\int_0^{t_d} h_r^2(t)dt}{\int_{t_d}^{\infty} h_r^2(t)dt}, \label{eq:DRR}
\end{align}
where the direct-path window is set to $t_d = \SI{2.5}{ms}$ following \cite{kuttruff2016room}. These parameters are computed per octave band (\SI{125}{Hz} to \SI{8}{kHz}) and as broadband averages, and probed alongside speaker identity as a complementary categorical factor.

Classical disentanglement metrics such as DCI~\cite{eastwood2018framework} operate at the dimension level, scoring how individual latent dimensions map to individual generative factors; in partition-level architectures, this mapping does not apply. Instead, we train a lightweight probe $f_{a,p}$ on each (attribute, partition) pair and measure performance $\mathcal{M}(a,p)$ with a task-appropriate metric (Pearson~$\rho$ for room parameters, top-1 accuracy for speaker identity). For factor~$k$ with intended partition $p^{\textrm{in}}$ and unintended partition $p^{\textrm{un}}$, the disentanglement gap
\begin{equation}
  \Delta_k = \mathcal{M}(k,p^{\textrm{in}}) - \mathcal{M}(k,p^{\textrm{un}})
  \label{eq:gap}
\end{equation}
quantifies how well factor~$k$ is confined to its intended partition. We write $\Delta_{\text{acc}}$ and $\Delta_{\rho}$ for the gap evaluated with top-1 accuracy and Pearson~$\rho$, respectively. Gaps are not comparable across factors, but each independently measures the degree of separation. This adapts DCI's informativeness score from dimensions to partitions, following established practice in speech~\cite{hsu2019disentangling, qian2022contentvec} and recently formalized as a general probe-based evaluation framework by Plachouras et al.~\cite{plachouras2025unified}. Because our probes are deliberately simple (see Sec.~\ref{sec:method}), the measured leakage constitutes a lower bound on actual information content.

\section{Proposed Probing Method}
\label{sec:method}

\subsection{Probing Framework}

We estimate $\Delta_{\text{acc}}$ and $\Delta_{\rho}$ by treating the pre-trained encoder of the \ac{AT} models from \cite{grundhuber2025acoustic} as a fixed feature extractor and training lightweight probes on each embedding partition independently. The encoder is an EnCodec-based \cite{defossez2022highfi} \ac{NAC} operating at \SI{16}{kHz} with a hop length of 320~samples. It produces 128-dimensional outputs split equally into a 64-dimensional speech partition $\mathtt{s}_{c,r}$ and a 64-dimensional acoustic partition $\mathtt{h}_{c,r}$, each quantized by an independent \ac{RVQ}. We evaluated all model configurations from \cite{grundhuber2025acoustic}, spanning different training task sets, quantization levels ($N \in \{4, 8, 16\}$ and unquantized), and temporal downsampling factors (1 to 120) for the acoustic partition. The downsampling is applied to $\mathtt{h}_{c,r}$ after the encoder by an integer factor, reducing its temporal resolution while the speech partition remains at full resolution, which lowers the effective bitrate of the acoustic stream.

\subsection{Architecture}

We deliberately use a simple \ac{MLP} so that probe accuracy reflects the information content of the embeddings rather than estimator capacity. For both probes, the partition embedding is mean-pooled over the temporal dimension to a fixed 64-dimensional vector, followed by three fully connected layers (one for the speaker identification probe) with 128 hidden units, ReLU activations, and 10\% dropout. Mean pooling handles variable temporal lengths arising from different downsampling factors and compresses non-stationary temporal variability in speech content, since room-acoustics and speaker identity are stationary per utterance. Both room-acoustic parameters and speaker identity are therefore predicted at the utterance level.

The regression probe predicts $\hat{\boldsymbol{\theta}} = [\widehat{T}_{60}, \widehat{C}_{50}, \widehat{\text{DRR}}]^\top$ via eight linear heads per acoustic parameter (i.e., seven per-octave-band (\SI{125}{Hz} to \SI{8}{kHz}) and one broadband) trained with a mean-squared-error loss. The classification probe predicts speaker identity via a $K$-way softmax head trained with cross-entropy loss. The total architecture has \SI{44440}{} parameters for room parameter estimation and \SI{21220}{} parameters for speaker identification.

\subsection{Dataset and Training}

Speech data was drawn from DNS5 \textit{read\_speech} \cite{dubey2023icassp}, assumed to be anechoic. \acp{RIR} were sourced from GWAsmall \cite{tang2022gwa}, preprocessed by truncating samples before the first absolute peak, normalized by maximum absolute value, and scaled by 0.25 to prevent clipping. $T_{60}$ values were taken from GWAsmall metadata; $C_{50}$ and \ac{DRR} were computed from the normalized \ac{RIR} via (\ref{eq:C50}) and~(\ref{eq:DRR}). Anechoic samples (10\% of the dataset) were assigned $T_{60} = $ \SI{0}{s}, $C_{50} = $ \SI{100}{dB}, $\text{DRR} = $ \SI{70}{dB}. These maxima serve as upper bounds on the values computed in the dataset and enable regression. Reverberant speech was generated by convolving \SI{3}{s} excerpts with one short ($T_{60} < $ \SI{0.25}{s}), and one long (\SI{0.4}{s} $ < T_{60} < $ \SI{1.2}{s}) \ac{RIR} drawn uniformly from the split's room pool and normalizing to the value range of $\pm 1$.

The full dataset comprised \num{60000} training samples (\SI{50}{h}), and \num{6000} samples each for validation and test. For room parameter estimation, speakers and rooms are mutually exclusive across splits. For the speaker identity probe, the top $K = 100$ speakers by utterance count are retained, split 80/10/10\% per speaker. Here, the dataset consists of \num{7806} training, 983 validation, and 983 test samples.

Both probes were trained up to 100 epochs using the AdamW optimizer with a learning rate of $10^{-3}$ and a weight decay of $10^{-4}$; the learning rate was reduced by a factor of~\num{2} after \num{7} epochs without validation-loss improvement, and early stopping was used with a patience of 15. Weighted random sampling (inverse class frequency) mitigates speaker imbalance in the classification probe.

\section{Evaluation}
\label{sec:eval}

\begin{figure*}[ht]
    \centering
    \includegraphics[width=\textwidth]{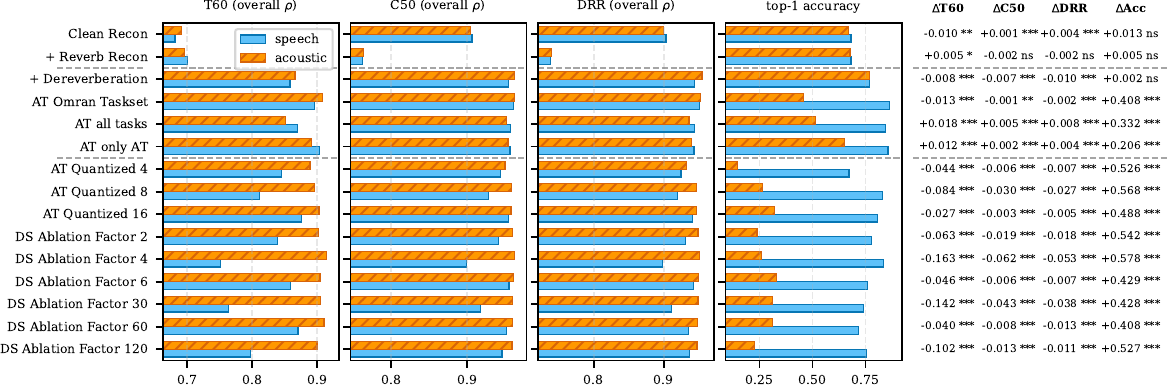}
    \caption{Overview of probe performance across four tasks: $T_{60}$, $C_{50}$, DRR (Pearson $\rho$ averaged over all bands and broadband), and speaker classification top-1 accuracy. Each row corresponds to one model; each panel shows paired bars for the speech (blue) and acoustic (orange) partitions. The rightmost table reports the per-model gap $\Delta$ with statistical significance annotations. Dashed separators indicate model family boundaries.}
    \label{fig:overview}
\end{figure*}

\begin{figure*}[ht]
    \centering \vspace{1em}
    \includegraphics[width=\textwidth]{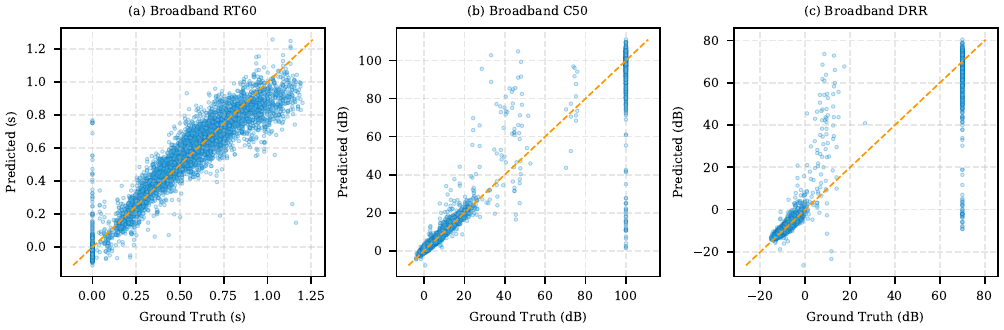}
    \caption{Predicted versus ground-truth broadband room-acoustic parameters from the acoustic embedding of model DS~Ablation Factor~4: (a)~$T_{60}$, (b)~$C_{50}$, and (c)~DRR. The dashed diagonal indicates the ideal prediction.}
    \label{fig:RPE_scatter_plots}
\end{figure*}

\subsection{Disentanglement}
Disentanglement was evaluated by comparing the gaps $\Delta_{\text{acc}}$ and $\Delta_{\rho}$.
Significance of $\Delta_\rho$ was assessed via the Steiger test~\cite{steiger1980tests} for dependent correlations, which accounts for the shared test set and ground-truth labels; $\Delta_{\text{acc}}$ was tested with a two-proportion $z$-test. Significance thresholds follow $p \leq 0.05$ (*), $p \leq 0.01$ (**), $p \leq 0.001$ (***), and $p > 0.05$ (ns). In Fig.~\ref{fig:overview}, we show the probe performance across four tasks: $T_{60}$, $C_{50}$, DRR (Pearson $\rho$ averaged over all bands and broadband), and speaker classification top-1 accuracy. Non-AT baselines (Clean Recon, +Reverb Recon, +Dereverberation) show very small speaker-identity gaps, none are statistically significant. All AT-trained models, by contrast, yield statistically significant separation on every probed factor. Speaker identity is effectively confined to the speech partition, with $\Delta_{\text{acc}}$ reaching 56.8~pp for the quantized $N\!=\!8$ model (83.1\% vs.\ 26.3\%), and the AT Omran Taskset reaches 86.5\% vs.\ 45.7\% speaker accuracy from the speech and acoustic partitions, respectively. \ac{AT} training appears required for the observed speaker separation.

Acoustic information, however, separated only partially: although the acoustic partition consistently yields higher room-parameter correlations, the speech partition retains acoustic information across all configurations. While quantization amplifies speaker separation ($\Delta_{\text{acc}} = 56.8$~pp at $N=8$ vs.\ 40.8~pp unquantized) by discarding entangled speaker residuals from the acoustic partition, speech-partition $T_{60}$ leakage remains above $\rho = 0.75$.
Temporal downsampling of the acoustic partition left room parameter estimation performance largely unchanged ($T_{60}$ $\rho \in [0.895, 0.915]$). This confirms the temporal compressibility of room-acoustic properties. For the downsampled acoustic embeddings, speech-partition leakage varies between $\rho = 0.752$ and $0.870$ without a systematic trend.

\subsection{Physically Meaningful Acoustic Embeddings}
\label{sec:physical_structure}

Despite receiving no room-parameter labels during codec training, the acoustic embeddings support competitive blind estimation, which underlines that the embedding carries physically meaningful room structure, complementing the disentanglement gaps of Sec.~\ref{sec:eval}. The obtained room-parameter estimates emerge as a by-product of a general-purpose codec embedding, without any task-specific feature engineering or room-parameter supervision. The best configuration (DS~Ablation Factor~4) achieves broadband $T_{60}$ \ac{RMSE} = \SI{0.094}{s} ($\rho = 0.947$), $C_{50}$ $\rho = 0.964$, and DRR $\rho = 0.954$ from its acoustic partition alone (Fig.~\ref{fig:RPE_scatter_plots}). Figure~\ref{fig:RPE_scatter_plots} shows the predicted versus ground-truth broadband values for all three parameters, where predictions closely follow the ideal diagonal across the full range of $T_{60}$, $C_{50}$, and \ac{DRR}. Predictions for anechoic signals show large outliers where estimation errors for high $C_{50} >$ \SI{30}{dB} or high DRR $>$ \SI{30}{dB} are very large. 
Per-band performance follows the expected frequency dependence, weakest at \SI{125}{Hz} ($T_{60}$ $\rho = 0.800$) and strongest at \SI{8}{kHz} ($\rho = 0.940$).

\begin{table}[t]
\centering
\caption{Broadband $T_{60}$ estimation comparison to baselines.}
\label{tab:baselines}
{\setlength{\tabcolsep}{4pt} 

\begin{tabular}{lccc}
\hline
\textbf{Method} & \textbf{RMSE [s]} & \textbf{MAE [s]} & $\rho$ \\ \hline
Löllmann ML \cite{lollmann2010improved}   & 0.404 & 0.308 & 0.446 \\
Spectrogram CNN \cite{gamper2018blind}    & 0.087 & 0.064 & 0.955 \\
CRNN-MB \cite{gotz2023online}             & \textbf{0.082} & \textbf{0.056} & \textbf{0.959} \\
Log-mel MLP (ours)                   & 0.204 & 0.163 & 0.564 \\
Acoustic Emb.\ MLP (ours)                & 0.094 & 0.064 & 0.947 \\
\hline
\end{tabular}}
\vspace{-1em}
\end{table}

Table~\ref{tab:baselines} contextualizes these results against supervised baselines retrained on the same data.
The acoustic-embedding MLP falls within \SI{0.02}{s} RMSE of the fully supervised CRNN-MB (\SI{0.082}{s}, $\rho = 0.959$) and spectrogram CNN (\SI{0.087}{s}, $\rho = 0.955$), using only a 64-dimensional embedding generated using a pre-trained encoder.
A controlled ablation using the same MLP architecture on 64-dimensional log-mel features yields substantially worse performance (RMSE = \SI{0.204}{s}, $\rho = 0.564$), confirming that the codec embedding contributes to estimation quality.

\section{Discussion}
\label{sec:discussion}

The asymmetry between speech and acoustic embedding leakage can be attributed back to the AT training objective: Embedding swaps across speakers force the decoder to recover speaker identity exclusively from the speech partition. This directly penalizes speaker leakage into the acoustic embedding. No analogous task minimizes room-acoustic information in the speech partition. If the speech embedding redundantly encodes room characteristics, the decoder ignores them when a different acoustic embedding is supplied.

Importantly, output-quality metrics are blind to these distinctions. Comparing the disentanglement analysis to quality estimates in \cite{grundhuber2025acoustic} the +Dereverberation baseline achieves competitive ScoreQ without any detectable disentanglement, while the lower-quality AT-only model (ScoreQ~NR~2.95) yields higher probe informativeness ($\rho = 0.89$). The cross-reconstruction metric used in prior work conflates the two disentanglement axes. Probing is required to expose leakage that output metrics cannot detect.

This work extended the probing framework for disentanglement evaluation in \acp{NAC} by incorporating a regression task. This shows that, within the \ac{AT} training framework, speaker identity is effectively separated, whereas acoustic information continues to leak into the speech partition. 

No existing task penalizes room-acoustic information in the speech embedding. Future work could extend the training task to explicitly push acoustic information out of the speech embedding, e.g., by adversarial decorrelation via a gradient-reversal branch that predicts room parameters from $\mathtt{s}$. This would explicitly penalize the acoustic information that the speech encoder currently retains without cost. 

The present study probes only time-invariant factors (room-acoustics, speaker identity). Extending the framework to linguistic content, e.g., with phoneme or ASR-based probes on both partitions, would establish a third disentanglement and could verify whether the non-downsampled acoustic partition is truly speech-free. Evaluating the framework on codec architectures beyond EnCodec would further test the generality of the observed asymmetry. The \ac{MLP} probes are deliberately simple. Therefore, the reported leakage levels constitute a lower bound, and actual information leakage may exceed what is measured. Additionally, exploring asymmetric partition dimensionalities may offer a mechanism to shift the leakage balance.

\section{Conclusion}
\label{sec:conclusion}
We introduced a probing-based framework that quantifies disentanglement in \acp{NAC} by training lightweight regressors and classifiers on fixed embedding partitions, thereby directly measuring the information content per partition and per factor. Applying this framework to an \ac{AT} codec reveals an asymmetric disentanglement structure: speaker identity is effectively confined to the speech partition, while acoustic information leaks into speech partition embeddings. We trace this asymmetry to the \ac{AT} objective, which induces a direct optimization signal to remove speaker information from the acoustic partition but provides no analogous signal to push room information out of the speech partition. Quantization amplifies speaker separation but leaves acoustic leakage unchanged. The acoustic embeddings nevertheless encode physically meaningful room structure, achieving $T_{60}$ estimation with an RMSE of \SI{0.094}{s} within \SI{0.02}{s} of supervised baselines. We showed that cross-reconstruction evaluation fails to expose leakage, making probing necessary, and explicit decorrelation losses are needed for complete separation along both axes.


\section{Acknowledgements}
The authors gratefully acknowledge the scientific support and HPC resources provided by the Erlangen National High Performance Computing Center (NHR@FAU) of the Friedrich-Alexander-Universität Erlangen-Nürnberg (FAU). The hardware is funded by the German Research Foundation (DFG).

\section{Generative AI Use Disclosure}
Generative AI in the form of LLMs was used for \LaTeX~formatting, plotting, and minor script corrections.

\bibliographystyle{IEEEtran}
\bibliography{mybib}

\end{document}